\documentclass[preprint,showpacs,preprintnumbers,amsmath,amssymb]{revtex4}
\usepackage{graphicx}
\usepackage{dcolumn}
\usepackage{bm}
\usepackage{color}
\usepackage{hhline}

\begin{document}

\title{Prediction of the bias voltage dependent magnetic contrast in spin-polarized scanning tunneling microscopy}

\author{Kriszti\'an Palot\'as}
\email{palotas@phy.bme.hu}
\affiliation{Budapest University of Technology and Economics,
Department of Theoretical Physics, Budafoki \'ut 8., H-1111 Budapest, Hungary}

\date{\today}

\begin{abstract}

This work is concerned with the theoretical description of the contrast, i.e., the apparent height difference between
two lateral surface positions on constant current spin-polarized scanning tunneling microscopy (SP-STM) images.
We propose a method to predict the bias voltage dependent magnetic contrast from single point tunneling current or
differential conductance measurements, without the need of scanning large areas of the surface.
Depending on the number of single point measurements, the bias positions of magnetic contrast reversals and of the maximally
achievable magnetic contrast can be determined. We validate this proposal by
simulating SP-STM images on a complex magnetic surface employing a recently developed approach based on atomic superposition.
Furthermore, we show evidence that the tip electronic structure and magnetic orientation have a major effect on the magnetic
contrast. Our theoretical prediction is expected to inspire experimentalists to considerably reduce measurement efforts for
determining the bias dependent magnetic contrast on magnetic surfaces.

\end{abstract}


\maketitle

\section{Introduction}
\label{sec_intro}

Achieving ultrahigh information density in a controlled way on surfaces of materials \cite{ultrahigh,weiss05} is one of the
ultimate goals of magnetic research nowadays for the purpose of future data storage technological applications. This can be
established by the reduction of the size of magnetic information storage units going down to the nanoscale or even to single atoms
\cite{serrate10}. Reading and writing information routinely from and to such magnetic units is a great challenge.
Spin-polarized scanning tunneling microscopy (SP-STM) \cite{bode03review} employing a magnetic tip proved to be extremely
successful for studying magnetism on surfaces in high spatial resolution. Recent experimental advances using this technique allow
the investigation of complex magnetic structures (frustrated antiferromagnets, spin spirals, skyrmion lattices, etc.)
\cite{wiesendanger09review,wulfhekel10review,serrate10,heinze11skyrmion}.

In the most routinely used constant current mode of the SP-STM the apparent height difference between differently magnetized
surface atoms allows the discrimination of the individual atomic magnetic properties and the mapping of the magnetic structure.
This apparent height difference is called the magnetic contrast. Finding the maximal magnetic contrast for a given surface-tip
combination is crucial for a more efficient magnetic mapping. This can be done by using magnetic tips with large
spin polarization, or by choosing the appropriate bias voltage.
A very few works focused on the investigation of the bias dependent magnetic contrast so far.
Among those, a magnetic contrast reversal was reported in two different magnetic systems \cite{yang06,palotas11stm}, and the
effect was related to the surface electronic structure rather than to the effect of the tip. In another work such contrast
reversals were observed during the scanning with the STM tip at fixed bias, and this was identified to be due to the magnetic
switching of the tip \cite{wasniowska10}. Moreover, under certain circumstances, a giant magnetic contrast can be obtained, and
this effect was explained by chemically modified STM tips \cite{hofer08tipH}.

Distinguishing atoms with different magnetic properties on a complex magnetic surface can successfully be performed by
spin-polarized scanning tunneling spectroscopy (SP-STS) as well. For example this technique has recently been used to read out
information from an atomic scale all-spin-based logic device \cite{khajetoorians11logic}.
Here, through scanning the surface with a magnetic tip, the measured differential conductance ($dI/dV$) values vary depending on
the magnetic properties of the underlying surface atom. For such a spectroscopic detection of atomic magnetism, the contrast,
i.e., the $dI/dV$ difference above individual atoms, should also be tunable by changing the bias voltage \cite{palotas11sts}.

In this work we propose a method to predict the bias voltage dependent apparent height difference between two lateral surface
positions on constant current SP-STM images without the need of scanning the full surface magnetic unit cell,
but using single point tunneling current data or differential conductance spectra. Therefore,
a reconsideration of the relation between constant current and constant height STM images \cite{chen93} in the SP-STM scenario
is necessary, and we introduce two magnetic contrast formulas that contain quantities above the two lateral sites only.
Depending on the number of single point measurements with oppositely aligned tip magnetizations
different information on the magnetic contrast can be obtained:
Taking tunneling current/spectra at one single point, the bias position of the contrast reversal can be identified.
Measurements above two inequivalent lateral surface positions $A$ and $B$ at two different tip-sample distances
$z_1$ and $z_2$ in the combination of $(B,z_1)$, $(A,z_1)$, and $(A,z_2)$ (three points)
together with one of the contrast formulas enable the determination of the
bias dependent magnetic contrast between the given surface positions at the equivalent tip-sample distance $z_1$.
From this curve the bias position of the maximally achievable magnetic contrast can be obtained.
Employing the other contrast formula requires the recording of the tunneling current/spectra at an extra tip position $(B,z_2)$
(altogether four points). In addition, measurements at the specified four tip positions enable the determination of the bias
dependent magnetic contrast between $A$ and $B$ surface sites for arbitrary tip-sample separations assuming an exponential decay
for the magnitude of the contrast with respect to the tip-sample distance. We demonstrate the predictive capabilities of this
method above a complex magnetic surface by performing numerical simulations based on first principles electronic structure data
employing a recently developed atom superposition approach \cite{palotas11stm,palotas12sts}.
Comparing the bias dependence of the predicted magnetic contrasts to that of extracted from constant current SP-STM images we find
excellent agreement, and based on that we propose this approach to be applied to SP-STM/STS experimental data as well.
Moreover, we analyze the tip-sample distance dependence of the contrast, and also
show evidence that the tip electronic structure and magnetic orientation have a major effect on the magnetic contrast.

The paper is organized as follows: The reconsideration of the relation between constant current and constant height STM images
in the SP-STM setup together with the two proposed magnetic contrast formulas are reported in section \ref{sec_contr}.
The procedure for obtaining different levels of information on the bias dependent magnetic contrast from
single point tunneling current data or $dI/dV$ spectra is presented in section \ref{sec_spec}.
We validate our proposal by means of numerical simulations taking the complex magnetic surface
of one monolayer (ML) Cr on a Ag(111) substrate and two tip models with different spin polarization characters. Details of the
employed atom superposition approach are given in section \ref{sec_comp}, and the results are presented in section \ref{sec_res}.
The summary of our findings is found in section \ref{sec_concl}. Finally, in the appendix we report the derivation of the formula
for determining the bias dependent magnetic contrast for arbitrary tip-sample separations, Eq.(\ref{Eq_corrug4points}).

\section{Method}

\subsection{Contrast in SP-STM}
\label{sec_contr}

We define the contrast between two lateral surface positions as their apparent height difference on a constant current contour.

On a nonmagnetic constant current STM image the contrast between atoms $A$ and $B$ on the surface at the
average tip-sample distance $z_1$ and bias voltage $V$ is \cite{chen93,heinze06}
\begin{equation}
\label{Eq_DZ-nonmagn}
\Delta z^{AB}_{nonmagn}(z_1,V)=-\frac{\Delta I^{AB}(z_1,V)}{\partial I^{av}/\partial z(z_1,V)},
\end{equation}
where $\Delta I^{AB}(z_1,V)$ is the current difference above atoms $A$ and $B$ at the tip-sample distance $z_1$, and
$I^{av}(z,V)$ is a laterally averaged current over the surface chemical unit cell at a tip-sample distance of $z$.
Since it is a tunneling current, it decays exponentially as $z$ increases \cite{chen93}.

We adopt the above relation between constant current and constant height STM images to the SP-STM scenario.
In this case the total tunneling current $I_{TOT}$ can be written as the sum of a non-spin-polarized (topographic) part,
$I_{TOPO}$, and a spin-polarized (magnetic) part, $I_{MAGN}$, \cite{palotas11stm,heinze06,wortmann01}
\begin{equation}
\label{Eq_ITOT}
I_{TOT}=I_{TOPO}+I_{MAGN},
\end{equation}
and the total contrast can also be decomposed as the sum of topographic and magnetic contributions,
\begin{eqnarray}
\Delta z^{AB}(z_1,V)&=&\Delta z_{TOPO}^{AB}(z_1,V)+\Delta z_{MAGN}^{AB}(z_1,V)\nonumber\\
&=&-\frac{\Delta I_{TOPO}^{AB}(z_1,V)}{\partial I_{TOT}^{av}/\partial z(z_1,V)}-\frac{\Delta I_{MAGN}^{AB}(z_1,V)}{\partial I_{TOT}^{av}/\partial z(z_1,V)}.
\label{Eq_contrast}
\end{eqnarray}
Here, $\Delta I_{TOPO}^{AB}$ and $\Delta I_{MAGN}^{AB}$ are the respective topographic and magnetic current differences above
atoms $A$ and $B$. $I_{TOT}^{av}$ has to be calculated by laterally averaging the total current
over the surface (chemical and magnetic) supercell at a constant tip-sample distance,
\begin{equation}
\label{Eq_Iav}
I_{TOT}^{av}(z,V)=\frac{1}{N_x N_y}\sum_{i=1}^{N_x}\sum_{j=1}^{N_y}I_{TOT}(x_i,y_j,z,V),
\end{equation}
where $N_x$ and $N_y$ denote the number of grid points in the lateral $x$ and $y$ directions, respectively.
Again, $I_{TOT}^{av}(z,V)$ is expected to decay exponentially as $z$ increases.

In the following, let us focus on the magnetic contrast only. Therefore, we assume an atomically flat sample surface
consisting of chemically equivalent but magnetically inequivalent atoms. In this case the topographic contrast between
any two surface atoms disappears since $\Delta I_{TOPO}^{AB}=0$. Hence, the total contrast between surface atoms
is the magnetic contrast, $\Delta z^{AB}(z_1,V)=\Delta z_{MAGN}^{AB}(z_1,V)$.

Since the calculation of the contrast requires the $z$-derivative of the exponentially decaying laterally averaged total current
in Eq.(\ref{Eq_Iav}), a full scanning of the surface magnetic unit cell at two constant heights is necessary.
The measurement time of this is comparable to record the constant current contour above the same scan area, thus there is
no advantage of using Eq.(\ref{Eq_contrast}) for the contrast estimation.
We would like to avoid any scanning above the surface but still predict the magnetic contrast between two surface atoms
on a constant current contour. Therefore, the denominator in Eq.(\ref{Eq_contrast}) needs to be reconsidered, and it is allowed
to contain current quantities above the two lateral sites $A$ and $B$ only.

A motivation for constructing the magnetic contrast formula is suggested by the following analogy at constant current condition:
In a nonmagnetic STM image the modulation due to the surface atoms is superimposed on the average tip-sample distance, whereas
in an SP-STM image of a complex magnetic surface the magnetic modulation is superimposed on the topographic image.
Therefore, taking Eq.(\ref{Eq_DZ-nonmagn}) and generalizing to the SP-STM case, the small lateral variation of the current due to
the magnetic modulation plays the role of the numerator, i.e., $\Delta I^{AB}\rightarrow\Delta I_{MAGN}^{AB}$, and
the topographic current takes the role of the average current in the denominator, i.e., $I^{av}\rightarrow I_{TOPO}$.
This is a fortunate choice since the topographic currents above all surface atoms are the same due to the assumed
chemical equivalence, $I_{TOPO}^A(z,V)=I_{TOPO}^B(z,V)$.
Following this, we can define a magnetic contrast between atoms $A$ and $B$ on the surface at the tip-sample distance $z_1$
and bias voltage $V$ as
\begin{equation}
\label{Eq_DZ}
\Delta z^{AB}_I(z_1,V)=-\frac{\Delta I_{MAGN}^{AB}(z_1,V)}{\partial I_{TOPO}^A/\partial z(z_1,V)}=\frac{I_{MAGN}^A(z_1,V)-I_{MAGN}^B(z_1,V)}{2\kappa_{TOPO}^A(V)I_{TOPO}^A(z_1,V)},
\end{equation}
where we took advantage of the exponentially decaying character of the topographic tunneling current,
\begin{equation}
\label{Eq_Itopo}
I_{TOPO}^A(z,V)=I_{TOPO}^{A0}(V)e^{-2\kappa_{TOPO}^A(V)z}.
\end{equation}
We investigate the validity of this equation in section \ref{sec_cr1cr3}.

An even more straightforward idea is to approximate the lateral average of the total tunneling current over the magnetic unit cell
as the average of the currents measured above the $A$ and $B$ sites,
which is still supposed to decay exponentially with respect to $z$ (for validation, see section \ref{sec_cr1cr3}),
\begin{equation}
\label{Eq_IavAB}
I_{TOT}^{av.AB}(z,V)=\frac{I_{TOT}^A(z,V)+I_{TOT}^B(z,V)}{2}=I_{TOT}^{av.AB0}(V)e^{-2\kappa_{TOT}^{av.AB}(V)z}.
\end{equation}
Using this quantity, another magnetic contrast between atoms $A$ and $B$ on the surface at the tip-sample distance $z_1$ and
bias voltage $V$ can be defined as
\begin{equation}
\label{Eq_DZ-TOT}
\Delta z^{AB}_{II}(z_1,V)=-\frac{\Delta I_{MAGN}^{AB}(z_1,V)}{\partial I_{TOT}^{av.AB}/\partial z(z_1,V)}=\frac{I_{MAGN}^A(z_1,V)-I_{MAGN}^B(z_1,V)}{2\kappa_{TOT}^{av.AB}(V)I_{TOT}^{av.AB}(z_1,V)}.
\end{equation}

Eq.(\ref{Eq_DZ}) and Eq.(\ref{Eq_DZ-TOT}) are the two key results of the present work for the bias dependent magnetic contrast
estimation. We validate them by simulating SP-STM images above the Cr/Ag(111) surface, and extracting apparent height
differences from constant current contours in section \ref{sec_cr1cr3}. In the following we consider how the ingredients for
Eq.(\ref{Eq_DZ}) and Eq.(\ref{Eq_DZ-TOT}) can be obtained in SP-STM/STS experiments, and we report a procedure,
which gives different levels of information on the bias dependent magnetic contrast from single point tunneling current data or
$dI/dV$ spectra measured with oppositely magnetized tips.

\subsection{Magnetic contrast information from single point quantities}
\label{sec_spec}

Let us assume that one can measure the bias dependence of the tunneling currents $I_P^A(z_1,V)$ and $I_{AP}^A(z_1,V)$ at the fixed
tip position $(A,z_1)$ (single point), where $A$ denotes a lateral surface site, $z_1$ is the tip-sample distance, and
$P$ and $AP$ denote parallel and antiparallel tip magnetization orientations, respectively, compared to a predefined direction.
From these data the spin-polarized contribution to the current in Eq.(\ref{Eq_ITOT}) can be obtained as
\begin{equation}
\label{Eq_I_MAGN}
I_{MAGN}^A(z_1,V)=\frac{I_P^A(z_1,V)-I_{AP}^A(z_1,V)}{2}.
\end{equation}
In the following we refer to this quantity as the magnetic current. If the measurement of differential conductances is
for any reason easier than of tunneling currents, then the magnetic current can still be obtained from
$dI_P^A/dV(z_1,\tilde{V})$ and $dI_{AP}^A/dV(z_1,\tilde{V})$, which have to be recorded at the same tip position $(A,z_1)$
and $P$ and $AP$ tip magnetization directions for a series of bias voltages $\tilde{V}$.
From these two series of tunneling spectra the spin-polarized contribution to the differential conductance can be calculated as
\begin{equation}
\label{Eq_didv_MAGN}
\frac{dI_{MAGN}^A}{dV}(z_1,\tilde{V})=\frac{1}{2}\left[\frac{dI_P^A}{dV}(z_1,\tilde{V})-\frac{dI_{AP}^A}{dV}(z_1,\tilde{V})\right].
\end{equation}
Here, we used the fact that the differential conductance can be decomposed as the sum of non-spin-polarized and spin-polarized
contributions \cite{palotas11sts,palotas12sts}, similarly to Eq.(\ref{Eq_ITOT}).
Thus, using $dI_{MAGN}^A/dV(z_1,\tilde{V})$ the magnetic current can be determined at arbitrary $V$ bias voltages via integration:
\begin{equation}
\label{Eq_Integral_MAGN-exp}
I_{MAGN}^A(z_1,V)=\int_{0}^{V}d\tilde{V}\frac{dI_{MAGN}^A}{dV}(z_1,\tilde{V}).
\end{equation}
Here, the integral limits correspond to zero temperature.

Assuming structural stability for the tip and the surface during the above described measurements we can determine the
bias voltage(s), where the magnetic current is zero, directly from Eq.(\ref{Eq_I_MAGN}),
or by varying the upper integral limit $V$ in Eq.(\ref{Eq_Integral_MAGN-exp}).
As demonstrated in section \ref{sec_res}, the zero magnetic current corresponds to a magnetic contrast inversion
on the constant current SP-STM image. Thus, from $P$ and $AP$ single point current data or $dI/dV$ spectra measured at the
tip position $(A,z_1)$ the bias voltage(s) can be determined, where a magnetic contrast reversal occurs.
This is the first level of information on the bias dependent magnetic contrast based on single point measurements.
We investigate the sensitivity of the bias position of the contrast reversal depending on the
magnetization direction and the position of the tip in section \ref{sec_cr1cr3}.

In order to quantify the magnetic contrast between two lateral surface positions $A$ and $B$,
Eq.(\ref{Eq_DZ}) or Eq.(\ref{Eq_DZ-TOT}) needs to be employed.
Using Eq.(\ref{Eq_DZ}), measurements at three tip positions are necessary: $(B,z_1)$, $(A,z_1)$, and $(A,z_2)$.
The first two positions are needed to obtain the numerator of Eq.(\ref{Eq_DZ}), i.e., the difference between $I_{MAGN}^A(z_1,V)$
and $I_{MAGN}^B(z_1,V)$. The magnetic current above surface site $B$, $I_{MAGN}^B(z_1,V)$, can similarly be calculated as
above surface site $A$, i.e., following the procedure reported at the beginning of this section,
Eq.(\ref{Eq_I_MAGN})-Eq.(\ref{Eq_Integral_MAGN-exp}).
For the denominator of Eq.(\ref{Eq_DZ}) the topographic current $I_{TOPO}^A(z_1,V)$, and its vacuum decay constant
$\kappa_{TOPO}^A(V)$ have to be determined from experiments. The topographic current can be expressed using the already available
$I_P^A(z_1,V)$ and $I_{AP}^A(z_1,V)$ current data at tip position $(A,z_1)$ as
\begin{equation}
\label{Eq_I_TOPO}
I_{TOPO}^A(z_1,V)=\frac{I_P^A(z_1,V)+I_{AP}^A(z_1,V)}{2},
\end{equation}
or from the available tunneling spectra $dI_P^A/dV(z_1,\tilde{V})$ and $dI_{AP}^A/dV(z_1,\tilde{V})$. From
these two series of tunneling spectra the non-spin-polarized contribution to the differential conductance can be calculated as
\begin{equation}
\label{Eq_didv_TOPO}
\frac{dI_{TOPO}^A}{dV}(z_1,\tilde{V})=\frac{1}{2}\left[\frac{dI_P^A}{dV}(z_1,\tilde{V})+\frac{dI_{AP}^A}{dV}(z_1,\tilde{V})\right].
\end{equation}
Using this quantity the topographic current can be determined at arbitrary $V$ bias voltages via integration:
\begin{equation}
\label{Eq_Integral_TOPO-exp}
I_{TOPO}^A(z_1,V)=\int_{0}^{V}d\tilde{V}\frac{dI_{TOPO}^A}{dV}(z_1,\tilde{V}).
\end{equation}
Again, the integral limits correspond to zero temperature.

In order to be able to calculate the vacuum decay constant $\kappa_{TOPO}^A(V)$ in Eq.(\ref{Eq_Itopo}), where an exponential decay
was assumed, the topographic current has to be obtained at a different tip-sample distance $z_2$ above site $A$: $(A,z_2)$.
Therefore, measurements of tunneling currents or $dI/dV$ spectra with oppositely magnetized tips at this tip position are
necessary. From these $I_{TOPO}^A(z_2,V)$ can similarly be determined as $I_{TOPO}^A(z_1,V)$, i.e., following the procedure
reported in the previous paragraph, Eq.(\ref{Eq_I_TOPO})-Eq.(\ref{Eq_Integral_TOPO-exp}). From expressing Eq.(\ref{Eq_Itopo}) at
tip-sample distances $z_1$ and $z_2$, the vacuum decay constant of the topographic current above atom $A$ can be given as
\begin{equation}
\label{Eq_kappatopo}
\kappa_{TOPO}^A(V)=\frac{\ln I_{TOPO}^A(z_1,V)-\ln I_{TOPO}^A(z_2,V)}{2(z_2-z_1)}.
\end{equation}
The derivation of this formula is identical to that of Eq.(\ref{Eq_kappadz}) in the appendix.
Note that though the absolute tip-sample distances $z_1$ and $z_2$ are unknown in experiments,
the tip displacement $z_2-z_1$ can be experimentally obtained.

Thus, by measuring single point current data or $dI/dV$ spectra with oppositely magnetized $P$ and $AP$ tips at three well-defined
positions above the surface: $(B,z_1)$, $(A,z_1)$, and $(A,z_2)$, the bias dependent magnetic contrast between atoms $A$ and $B$
at the tip-sample distance $z_1$, $\Delta z^{AB}_I(z_1,V)$, can be predicted following Eq.(\ref{Eq_DZ}). This is the second level
of information on the bias dependent magnetic contrast at a fixed tip-sample distance. From this function the bias position can be
identified, where the maximal magnetic contrast can be achieved. We investigate the sensitivity of this bias position
depending on the tip-sample distance and the magnetization direction of the tip in section \ref{sec_cr1cr3}.

We proposed a second option for the prediction of the bias dependent magnetic contrast in Eq.(\ref{Eq_DZ-TOT}).
For using this formula, tunneling current or $dI/dV$ data at four tip positions are needed: $(A,z_1)$, $(B,z_1)$,
$(A,z_2)$, and $(B,z_2)$. Measurements at the first two positions with oppositely magnetized tips are necessary to obtain the
numerator of Eq.(\ref{Eq_DZ-TOT}), which is the same as of Eq.(\ref{Eq_DZ}), i.e., the difference between $I_{MAGN}^A(z_1,V)$
and $I_{MAGN}^B(z_1,V)$. The magnetic currents above surface site $A$ and $B$ can be calculated from single point quantities
following the procedure reported at the beginning of this section, Eq.(\ref{Eq_I_MAGN})-Eq.(\ref{Eq_Integral_MAGN-exp}).
For the denominator of Eq.(\ref{Eq_DZ-TOT}) the total current averaged above the $A$ and $B$ sites, $I_{TOT}^{av.AB}(z_1,V)$ in
Eq.(\ref{Eq_IavAB}), and its vacuum decay constant $\kappa_{TOT}^{av.AB}(V)$ have to be determined. Therefore, the bias dependence
of the tunneling currents has to be measured at all of the mentioned four tip positions with a fixed $P$ tip magnetization
orientation:
$I_P^J(z_i,V)$, where $J\in\{A,B\}$ and $i\in\{1,2\}$. These data can also be obtained from tunneling spectra as
\begin{equation}
I_P^J(z_i,V)=\int_{0}^{V}d\tilde{V}\frac{dI_P^J}{dV}(z_i,\tilde{V}).
\end{equation}
The averaged currents at the two different tip-sample distances $z_1$ and $z_2$ are
\begin{equation}
\label{Eq_ITOTavAB}
I_{TOT}^{av.AB}(z_1,V)=\frac{I_P^A(z_1,V)+I_P^B(z_1,V)}{2},\;I_{TOT}^{av.AB}(z_2,V)=\frac{I_P^A(z_2,V)+I_P^B(z_2,V)}{2},
\end{equation}
respectively.
Using these and the assumed exponential decay in Eq.(\ref{Eq_IavAB}), the corresponding vacuum decay constant can be obtained as
\begin{equation}
\label{Eq_kappa_av.AB}
\kappa_{TOT}^{av.AB}(V)=\frac{\ln I_{TOT}^{av.AB}(z_1,V)-\ln I_{TOT}^{av.AB}(z_2,V)}{2(z_2-z_1)}.
\end{equation}
Thus, by measuring single point current data or $dI/dV$ spectra with oppositely magnetized $P$ and $AP$ tips at four well-defined
positions above the surface: $(A,z_1)$, $(B,z_1)$, $(A,z_2)$, and $(B,z_2)$, the bias dependent magnetic contrast between atoms
$A$ and $B$ at the tip-sample distance $z_1$, $\Delta z^{AB}_{II}(z_1,V)$, can be predicted following Eq.(\ref{Eq_DZ-TOT}).
We investigate the reliability of the magnetic contrast formulas Eq.(\ref{Eq_DZ}) and Eq.(\ref{Eq_DZ-TOT}) by explicitly
simulating SP-STM images, and extracting apparent height differences on a constant current contour between atoms of different
magnetic properties in section \ref{sec_cr1cr3}.

Finally, having single point current or $dI/dV$ data measured with oppositely magnetized tips at {\emph{all}} of the mentioned
four tip positions enables one to calculate the magnetic contrast at the tip-sample distance $z_2$, $\Delta z^{AB}(z_2,V)$,
as well, employing either Eq.(\ref{Eq_DZ}) or Eq.(\ref{Eq_DZ-TOT}).
Assuming an exponentially decaying magnitude of the contrast with increasing tip-sample distance, the bias dependent magnetic
contrast between atoms $A$ and $B$ at an {\emph{arbitrary}} tip-sample distance $z$ in the tunneling regime can be estimated as
\begin{equation}
\Delta z^{AB}(z,V)=sgn(\Delta z^{AB}(z_1,V))\times |\Delta z^{AB}(z_1,V)|^{\frac{z_2-z}{z_2-z_1}}\times |\Delta z^{AB}(z_2,V)|^{\frac{z-z_1}{z_2-z_1}}.
\label{Eq_corrug4points}
\end{equation}
This is the third level of information on the bias dependent magnetic contrast. The derivation of this formula is reported in the
appendix. We check the reliability of Eq.(\ref{Eq_corrug4points}) based on contrasts calculated by using Eq.(\ref{Eq_DZ})
in section \ref{sec_cr1cr3}.

The summary of the presented procedure to obtain different levels of information on the bias dependent magnetic contrast from
single point differential conductance or current measurements is given in Table \ref{Table_method}. In the remaining of the paper
we simulate the measurements following a simple model based on atomic superposition and first principles electronic structure data.

\subsection{Simulation: Atom superposition approach}
\label{sec_comp}

Recently, Palot\'as et al.\ developed a three-dimensional atom superposition approach for simulating SP-STM \cite{palotas11stm}
and SP-STS \cite{palotas12sts} on complex magnetic surfaces based on previous theories
\cite{heinze06,wortmann01,yang02,smith04,passoni09}. The model is inspired by the spin-polarized Tersoff-Hamann approach
\cite{wortmann01,tersoff83,tersoff85}, and assumes elastic tunneling through one tip apex atom.
The contributions from individual transitions between this apex atom and each surface atom are summed up assuming
the one-dimensional Wentzel-Kramers-Brillouin approximation in all these transitions.
The electronic structure of the sample and the tip is considered in the model by the projected electron density of states (PDOS)
of the tip apex and of the sample surface atoms obtained by {\it ab initio} electronic structure calculations.

Following this method, the topographic ($TOPO$) and magnetic ($MAGN$) components of the total tunneling current in
Eq.(\ref{Eq_ITOT}) at bias voltage $V$ and at the tip apex position $\underline{R}_{TIP}^{(J,z)}=(x_J,y_J,z)$ above surface site
$J$ can be determined as
\begin{eqnarray}
I_{TOPO}^J(z,V)
&\propto&\int_{0}^{V}dU\sum_{\alpha}n_S^{\alpha}(E_F^S+eU)n_T(E_F^T-eV+eU)
e^{-\frac{2}{\hbar}\sqrt{2m\left(\frac{\phi_S+\phi_T+eV}{2}-eU\right)}|\underline{R}_{TIP}^{(J,z)}-\underline{R}_{\alpha}|},\nonumber\\
\label{Eq_I_TOPO-atom_superpos}\\
I_{MAGN}^J(z,V)
&\propto&\int_{0}^{V}dU\sum_{\alpha}\underline{m}_S^{\alpha}(E_F^S+eU)\underline{m}_T(E_F^T-eV+eU)
e^{-\frac{2}{\hbar}\sqrt{2m\left(\frac{\phi_S+\phi_T+eV}{2}-eU\right)}|\underline{R}_{TIP}^{(J,z)}-\underline{R}_{\alpha}|}.\nonumber\\
\label{Eq_I_MAGN-atom_superpos}
\end{eqnarray}
These quantities at different tip positions are necessary to predict the various levels of information on the bias dependent
magnetic contrast (see Table \ref{Table_method}), as well as to simulate SP-STM images.
Here, the sum over $\alpha$ includes a sufficient number of surface atoms \cite{palotas11sts} with position vector
$\underline{R}_{\alpha}$, projected charge DOS $n_S^{\alpha}$, and magnetization DOS vector $\underline{m}_S^{\alpha}$
\cite{palotas11stm}. Similarly, $n_T$ and $\underline{m}_T$ denote the charge DOS and the magnetization DOS vector
projected onto the tip apex atom, respectively. $E_F^S$ and $E_F^T$ are the Fermi energies of the surface and the tip. Moreover,
the exponential expression is the tunneling transmission assuming spherical exponential decay of the electron wave functions
and an effective rectangular tunnel barrier, where $m$ is the electron mass, $\hbar$ is the reduced Planck constant,
and $\phi_S$ and $\phi_T$ are the surface and tip electron work functions, respectively. The effect of the electron orbitals
on the tunneling is neglected (independent orbital approximation \cite{heinze06}).
Note that a better description of the electron tunneling can be achieved by incorporating such orbital effects, e.g.,
by prescribing the tip orbital symmetry \cite{chen92,passoni07,donati11}, or by taking into account symmetry-decomposed
electronic structures and an orbital-dependent transmission function \cite{palotas12orb}.
Using an orbital-dependent tunneling model, contrast reversals of nonmagnetic origin are expected \cite{chen92,palotas12orb}.
In the present work we do not consider the topographic part of the contrast in Eq.(\ref{Eq_contrast}), and we focus on the
bias dependent magnetic contrast and its reversal only.

\section{Results and discussion}
\label{sec_res}

In order to demonstrate the applicability of the proposed procedure for predicting the bias dependent magnetic contrast from
single point tunneling data, we perform simulations on a sample surface with noncollinear magnetic order.
One monolayer Cr on Ag(111) is a prototype of frustrated hexagonal antiferromagnets \cite{heinze06}.
Due to the geometrical frustration of the antiferromagnetic exchange interactions between the Cr spin moments, its magnetic
ground state is a noncollinear $120^{\circ}$ N\'eel state \cite{wortmann01}. Taking the spin-orbit coupling into account,
two N\'eel states with opposite chiralities can form, and one of them is energetically favored \cite{palotas11stm}.

\begin{figure*}
\includegraphics[width=0.5\columnwidth,angle=0]{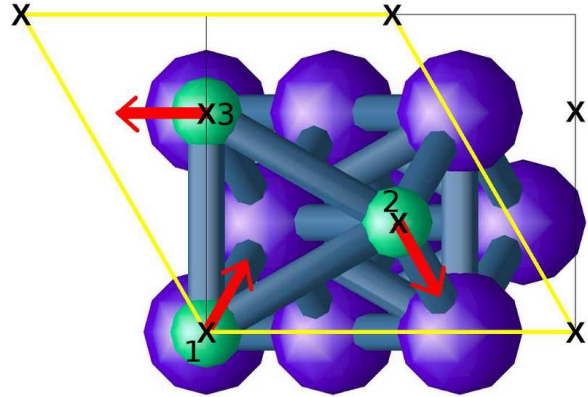}
\caption{\label{Fig1} (Color online) The surface geometry of 1 ML Cr on Ag(111) and its ground state magnetic structure.
The Cr and Ag atoms are denoted by spheres colored by green (medium gray) and purple (dark gray), respectively, and the
magnetic moments of individual Cr atoms are indicated by (red) arrows. The ($\sqrt{3}\times\sqrt{3}$) magnetic unit cell is drawn
by yellow (light gray) color, and the scanning area for the SP-STM simulations is shown by the black-framed rectangle. Moreover,
the surface Cr positions are denoted by "x", and Cr atoms in the magnetic unit cell are explicitly labeled by "1", "2", and "3".
}
\end{figure*}

We performed fully noncollinear electronic structure calculations on the Cr/Ag(111) system, based on the density functional theory
(DFT) within the generalized gradient approximation (GGA) and using the projector augmented wave (PAW) method, implemented in the
Vienna ab initio simulation package (VASP) \cite{VASP2,VASP3,kresse99,hobbs00prb,hobbs00jpcm}.
The computational details are reported elsewhere \cite{palotas11stm}.
The ground state magnetic structure is shown in Figure \ref{Fig1}, where we explicitly labeled the individual Cr atoms in the
($\sqrt{3}\times\sqrt{3}$) magnetic unit cell. The spin polarization of the surface Cr atoms is positive with respect to the
direction of the corresponding Cr magnetic moments below $E_F^S+0.54$ eV, and negative above this energy. More details on the
energy dependence of the Cr spin polarization can be found in Ref.\ \cite{palotas11stm}.

We used two tip models: One is an electronically flat,
maximally spin-polarized ($P_T=+1$) ideal magnetic tip, and the second is a blunt Ni tip, i.e., a Ni adatom placed above the
hollow position of a Ni(110) surface. The Ni tip apex atom has a high negative spin polarization close to the Fermi level,
i.e., $P_T=-0.91$ at $E_F^T$, and $|P_T(E)|>0.8$ between $E_F^T-0.3$ eV and $E_F^T+0.3$ eV.
More details on the energy dependence of the spin polarization of the Ni tip model can be found in Ref.\ \cite{palotas11sts}.

The charge and magnetization DOS of the surface atoms and the tip apex obtained from the above calculations were included in the
formulas Eq.(\ref{Eq_I_TOPO-atom_superpos}) and Eq.(\ref{Eq_I_MAGN-atom_superpos}) for determining the current components.
For the SP-STM images we calculated the total tunneling current in a box above the magnetic unit cell
containing 153000 ($34\times 30\times 150$) grid points with a 0.15 \AA\; lateral and 0.053 \AA\; horizontal resolution.

\subsection{Magnetic contrast reversal}
\label{sec_rev}

According to section \ref{sec_spec}, the first level of information on the bias dependent magnetic contrast is the
bias voltage of the contrast reversals, which can be obtained from the magnetic current at a single tip position.
Figure \ref{Fig2} shows the simulated magnetic currents $I_{MAGN}^J(z,V)$ according to Eq.(\ref{Eq_I_MAGN-atom_superpos})
at $z=3.5$ \AA\; above each Cr atom ($J\in\{$Cr1,Cr2,Cr3$\}$) in the ($\sqrt{3}\times\sqrt{3}$) magnetic unit cell
(see Figure \ref{Fig1}), measured with the two considered tip models. For the negative bias range the integral limits are reversed
in order to obtain positive total current values, and the magnetic current is calculated accordingly.
The tip magnetization direction is fixed parallel to the Cr1 magnetic moment. Due to the noncollinear magnetic structure, the Cr2
and Cr3 magnetic moments have an angle of $120^{\circ}$ with respect to the tip magnetization direction.
Following this, the magnetic currents measured above the Cr2 and Cr3 surface atoms are equal and their values are
$\cos(120^{\circ})=-0.5$ times the magnetic current above Cr1, i.e.,
\begin{equation}
\label{Eq_ICr3}
I_{MAGN}^{Cr2}=I_{MAGN}^{Cr3}=-I_{MAGN}^{Cr1}/2.
\end{equation}
Moreover, it is clearly demonstrated in Figure \ref{Fig2}
that the different tip spin polarization characters affect the magnetic current considerably. While the magnetic current above
Cr1 is positive and monotonically changing in the negative bias range for the ideal tip, it is negative and has a local minimum
for the Ni tip in the same bias range. At positive bias voltages the magnetic current is even more complicated, and it has
a local maximum for the ideal tip (see inset), whereas it has two local maxima and one local minimum (see inset) for the Ni tip.
In the studied bias range the global maximum and minimum is at -2.5 V and 2.5 V, respectively, for the ideal tip, and
at 2.32 V and -1.38 V, respectively, for the Ni tip.

The magnetic contrast is reversed at zero magnetic current.
In order to find out the corresponding bias positions for the two considered tips, we have to zoom in the region
[0 V, 1 V]. The indicated rectangular area is shown in the inset of Figure \ref{Fig2}. It is clearly seen that the sign change of
the magnetic current occurs at 0.94 V and 0.74 V for the ideal and the Ni tip, respectively. This means that
although the reversal is, in principle, due to the surface electronic structure \cite{yang06,palotas11stm},
the tip plays a crucial role as well, since its electronic structure modifies the bias position of the reversal. Note that the
determination of the above bias positions does not depend on whether the tip is placed above Cr1, Cr2, or Cr3 atoms. Moreover, we
find that the bias voltage of the contrast reversals is stable within $\pm 0.01$ V placing the tip above other surface positions.

\subsection{Magnetic contrast between Cr1 and Cr3}
\label{sec_cr1cr3}

In order to quantify the magnetic contrast between two surface sites at a given tip-sample distance,
Eq.(\ref{Eq_DZ}) or Eq.(\ref{Eq_DZ-TOT}) has to be employed. This is the second level of information on the bias dependent
magnetic contrast. Following Table \ref{Table_method}, we calculate all current ingredients for the magnetic contrast formulas
using Eq.(\ref{Eq_I_TOPO-atom_superpos}) and Eq.(\ref{Eq_I_MAGN-atom_superpos}) above Cr1 and Cr3 atoms.
For the determination of the vacuum decay constants $\kappa_{TOPO}^{Cr1}(V)$ and $\kappa_{TOT}^{av.Cr1-Cr3}(V)$ in the denominator
of Eq.(\ref{Eq_DZ}) and Eq.(\ref{Eq_DZ-TOT}), respectively, Eq.(\ref{Eq_kappatopo}) and Eq.(\ref{Eq_kappa_av.AB})
were proposed based on experimental data at two tip-sample distances $z_1$ and $z_2$.
To prove that the exponentially decaying character of the corresponding currents with respect to the tip-sample distance $z$ in
Eq.(\ref{Eq_Itopo}) and Eq.(\ref{Eq_IavAB}) is a valid assumption, we did not consider two tip-sample distances only,
but employed a series of ordinary least squares linear regressions on the $\ln I(z,V)$ data taking 150 $z$ values in the range
[0.01 \AA, 7.95 \AA] for 500 bias voltages in the [-2.5 V, 2.5 V] interval. For all considered bias voltages we obtained Pearson
product-moment correlation coefficients better than $r(V)=-0.9999$. These $r(V)$ values justify our exponentially decaying current
assumption in Eq.(\ref{Eq_Itopo}) and Eq.(\ref{Eq_IavAB}) within the atom superposition approach. Note that orbital-dependent
tunneling effects can modify this finding. The decay constants $\kappa_{TOPO}^{Cr1}(V)$ and $\kappa_{TOT}^{av.Cr1-Cr3}(V)$
were determined from the linear regressions.

\begin{figure*}
\includegraphics[width=1.0\columnwidth,angle=0]{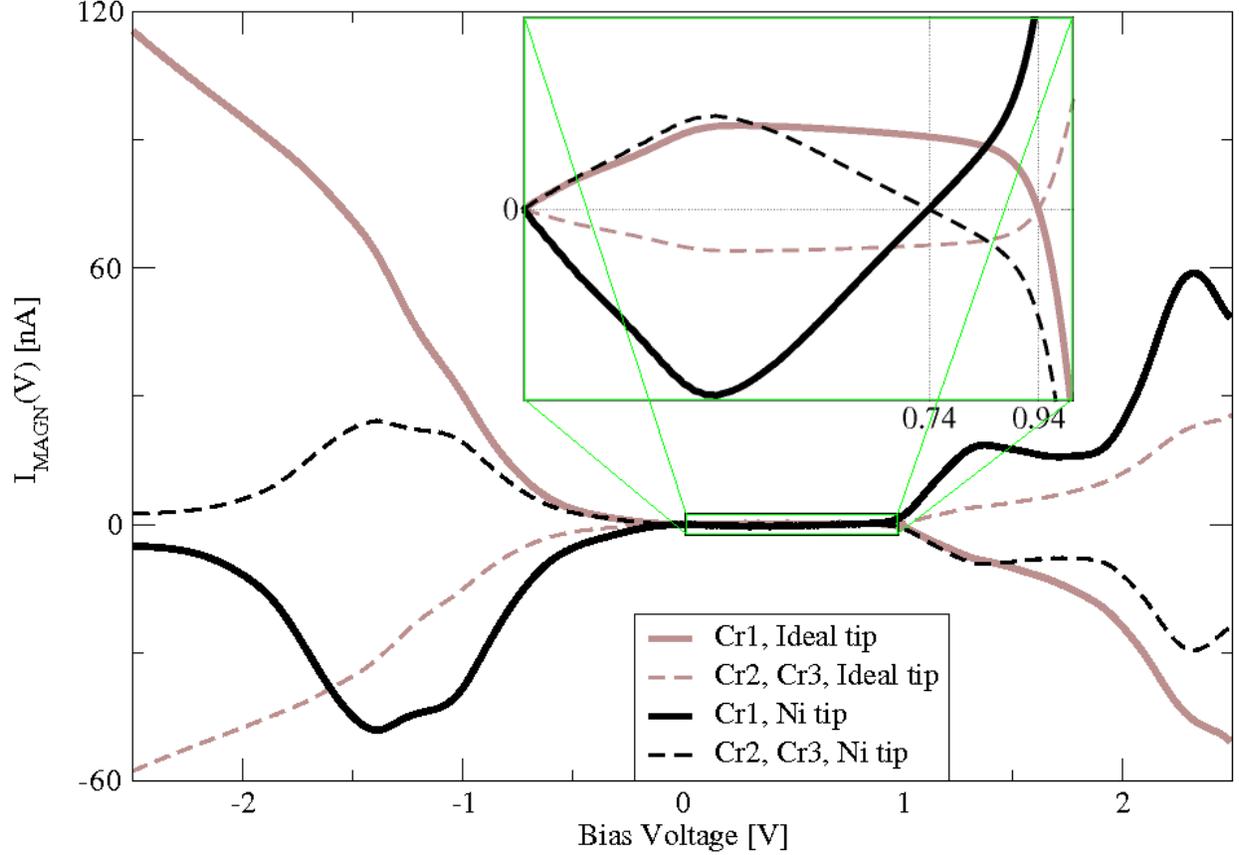}
\caption{\label{Fig2} (Color online) The simulated magnetic currents $I_{MAGN}^J(z=3.5$\AA$,V)$ according to
Eq.(\ref{Eq_I_MAGN-atom_superpos}) $z=3.5$ \AA\;above each Cr atom ($J\in\{$Cr1,Cr2,Cr3$\}$) in the ($\sqrt{3}\times\sqrt{3}$)
magnetic unit cell (see Figure \ref{Fig1}), measured with an ideal electronically flat maximally spin-polarized tip (gray),
or a model Ni tip (black). The tip magnetization direction is fixed parallel to the Cr1 magnetic moment.
The inset shows the bias region between 0 V and 1 V zoomed in. The sign change of the magnetic current occurs at 0.94 V and 0.74 V
for the ideal and the Ni tip, respectively. These correspond to magnetic contrast reversals.
}
\end{figure*}

\begin{figure*}
\includegraphics[width=1.0\columnwidth,angle=0]{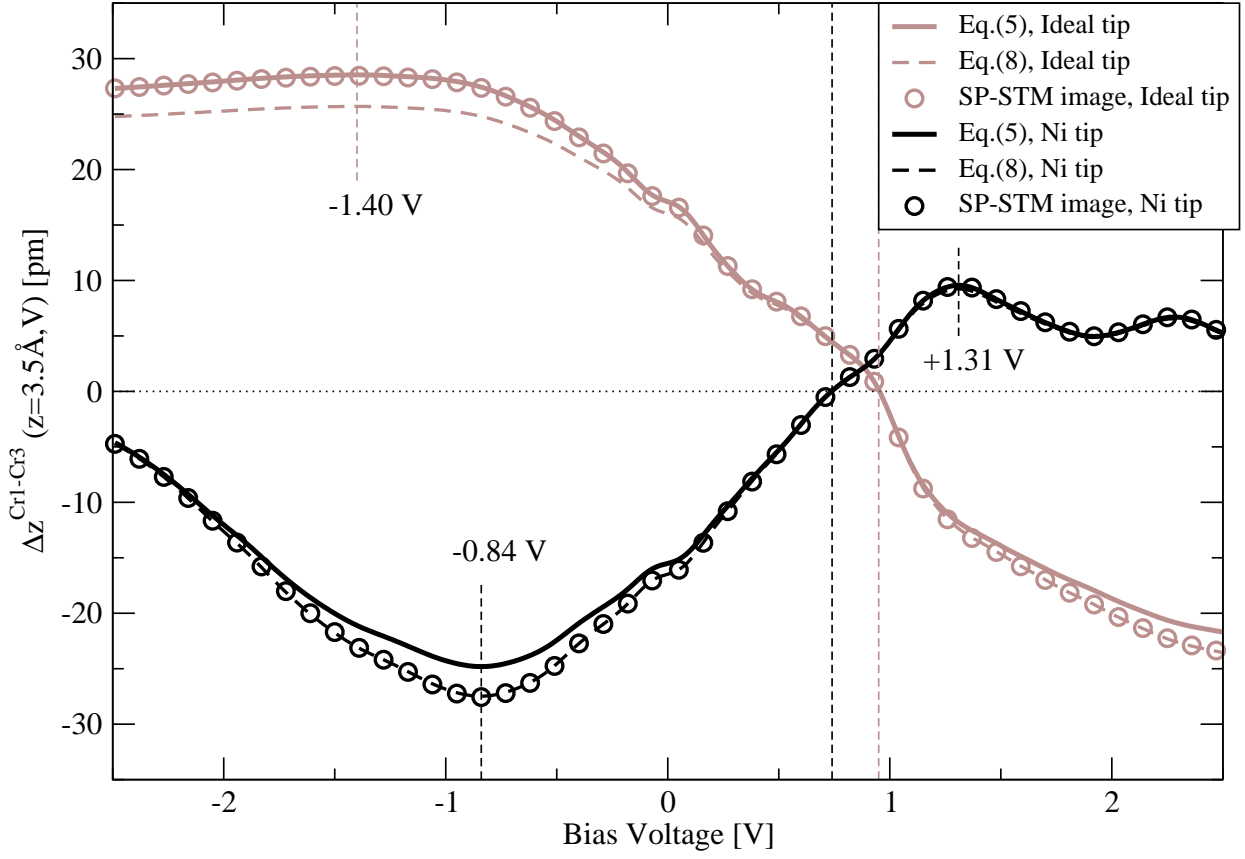}
\caption{\label{Fig3} (Color online) The bias dependent magnetic contrast between Cr1 and Cr3 atoms
$\Delta z^{Cr1-Cr3}(z=3.5$\AA$,V)$ at $z=3.5$ \AA\;tip-sample distance calculated using Eq.(\ref{Eq_DZ}) (solid lines) and
Eq.(\ref{Eq_DZ-TOT}) (dashed lines) measured with the ideal magnetic tip (gray) and the Ni tip (black).
The tip magnetization direction is fixed parallel to the Cr1 magnetic moment. The local absolute maxima
of the magnetic contrasts and their bias values are explicitly shown in both the negative and the positive bias ranges.
Vertical dashed lines denote the contrast reversals, see also Figure \ref{Fig2}.
For comparison, circle symbols show the apparent height difference between Cr1 and Cr3 atoms obtained from constant current
SP-STM images, see text for details.
}
\end{figure*}

Figure \ref{Fig3} shows the bias dependent magnetic contrast between Cr1 and Cr3 atoms at $z=3.5$ \AA\; tip-sample distance for
both contrast formulas Eq.(\ref{Eq_DZ}) and Eq.(\ref{Eq_DZ-TOT}) using the two tip models.
The tip magnetization direction is fixed parallel to the Cr1 magnetic moment. The sign of the curves corresponds to
those denoted by solid line in Figure \ref{Fig2}, $sgn(\Delta z^{Cr1-Cr3})=sgn(I_{MAGN}^{Cr1})$. This is understandable as the
sign of $\Delta z^{Cr1-Cr3}$ is determined by the sign of its numerator $\Delta I_{MAGN}^{Cr1-Cr3}$ since the denominator is
always positive, and $\Delta I_{MAGN}^{Cr1-Cr3}=(3/2)\times I_{MAGN}^{Cr1}$ because of Eq.(\ref{Eq_ICr3}). Accordingly,
the contrast reversal is obtained at $\Delta z^{Cr1-Cr3}=0$, i.e., at 0.94 V and 0.74 V for the ideal and the Ni tip,
respectively, in agreement with Figure \ref{Fig2}.
However, due to the denominator of Eq.(\ref{Eq_DZ}) and Eq.(\ref{Eq_DZ-TOT}), we obtain qualitatively different curves compared to
Figure \ref{Fig2}. From the calculated functions, bias voltages can be identified in both the negative and positive bias ranges,
where a local absolute maximum magnetic contrast can be achieved. As can be seen, these highly depend on the spin polarization
character of the magnetic tip. In the studied bias interval local absolute maximum contrasts are expected at
-1.40 V and 2.50 V for the measurement with the ideal tip, whereas these bias values are considerably modified to -0.84 V and
1.31 V by using the Ni tip. The maximum absolute contrasts are obtained at the corresponding negative bias values for both tips.
The reported bias positions do not depend on the employed magnetic contrast formula.

From Figure \ref{Fig3} it seems that the predicted magnetic contrast using Eq.(\ref{Eq_DZ-TOT}) (dashed lines) is always smaller
than using Eq.(\ref{Eq_DZ}) (solid lines). This means a larger absolute contrast for using Eq.(\ref{Eq_DZ-TOT}) if the contrast is
negative. Let us try to understand this difference. Therefore, we derive a relation between the two contrast formulas
$\Delta z^{Cr1-Cr3}_{I}$ in Eq.(\ref{Eq_DZ}) and $\Delta z^{Cr1-Cr3}_{II}$ in Eq.(\ref{Eq_DZ-TOT}). Using Eq.(\ref{Eq_ICr3})
and that $I_{TOPO}^{Cr3}=I_{TOPO}^{Cr1}$, the average total current according to Eq.(\ref{Eq_ITOTavAB}) is
\begin{equation}
I_{TOT}^{av.Cr1-Cr3}=\frac{(I_{TOPO}^{Cr1}+I_{MAGN}^{Cr1})+(I_{TOPO}^{Cr1}-I_{MAGN}^{Cr1}/2)}{2}=I_{TOPO}^{Cr1}+\frac{I_{MAGN}^{Cr1}}{4},
\end{equation}
and thus,
\begin{eqnarray}
\Delta z^{Cr1-Cr3}_{II}&=&\frac{\Delta I_{MAGN}^{Cr1-Cr3}}{2\kappa_{TOPO}^{Cr1}I_{TOPO}^{Cr1}+2\kappa_{MAGN}^{Cr1}I_{MAGN}^{Cr1}/4}
=\frac{\Delta I_{MAGN}^{Cr1-Cr3}}{2\kappa_{TOPO}^{Cr1}I_{TOPO}^{Cr1}}\times\frac{1}{1+\frac{1}{4}\frac{\kappa_{MAGN}^{Cr1}I_{MAGN}^{Cr1}}{\kappa_{TOPO}^{Cr1}I_{TOPO}^{Cr1}}}\nonumber\\
&=&\frac{\Delta z^{Cr1-Cr3}_{I}}{1+\frac{1}{4}\frac{\kappa_{MAGN}^{Cr1}\mathbf{I_{MAGN}^{Cr1}}}{\kappa_{TOPO}^{Cr1}I_{TOPO}^{Cr1}}},
\end{eqnarray}
where we omitted the $(z,V)$ or $(V)$ arguments of the quantities, and assumed an exponential decay for $I_{MAGN}^{Cr1}(z,V)$
with respect to the tip-sample distance. Since the quantities $\kappa_{MAGN}^{Cr1}$,
$\kappa_{TOPO}^{Cr1}$ and $I_{TOPO}^{Cr1}$ are always positive, the sign of $I_{MAGN}^{Cr1}$ determines the relation between
$\Delta z^{Cr1-Cr3}_{I}$ and $\Delta z^{Cr1-Cr3}_{II}$: If $I_{MAGN}^{Cr1}$ is positive then
$0<\Delta z^{Cr1-Cr3}_{II}<\Delta z^{Cr1-Cr3}_{I}$. On the other hand, if $I_{MAGN}^{Cr1}$ is negative then
$\Delta z^{Cr1-Cr3}_{II}<\Delta z^{Cr1-Cr3}_{I}<0$. This is exactly what we observe in Figure \ref{Fig3}.

For validating the bias dependent magnetic contrast predictions, we calculate the apparent height difference between Cr1 and Cr3
atoms from constant current SP-STM images, which are also simulated within the atom superposition approach. The current contours
are chosen in such a way that the apparent height of the Cr1 atom is 3.5 \AA\; at all considered bias voltages. The obtained data
for the two considered tip models are shown in Figure \ref{Fig3} using circle symbols. The qualitative agreement with the
predicted magnetic contrasts using Eq.(\ref{Eq_DZ}) and Eq.(\ref{Eq_DZ-TOT}) is obvious at the first sight. Having a closer look
we find that Eq.(\ref{Eq_DZ}) quantitatively reproduces the apparent height difference for $\Delta z^{Cr1-Cr3}>0$, and
Eq.(\ref{Eq_DZ-TOT}) for $\Delta z^{Cr1-Cr3}<0$. This means that both formulas are needed for a quantitative determination of the
bias dependent magnetic contrast. In case one is interested in the identification of the bias voltage for obtaining the maximum
contrast, the formula requiring less measurements or calculations, Eq.(\ref{Eq_DZ}), can be applied.

\begin{figure*}
\includegraphics[width=1.0\columnwidth,angle=0]{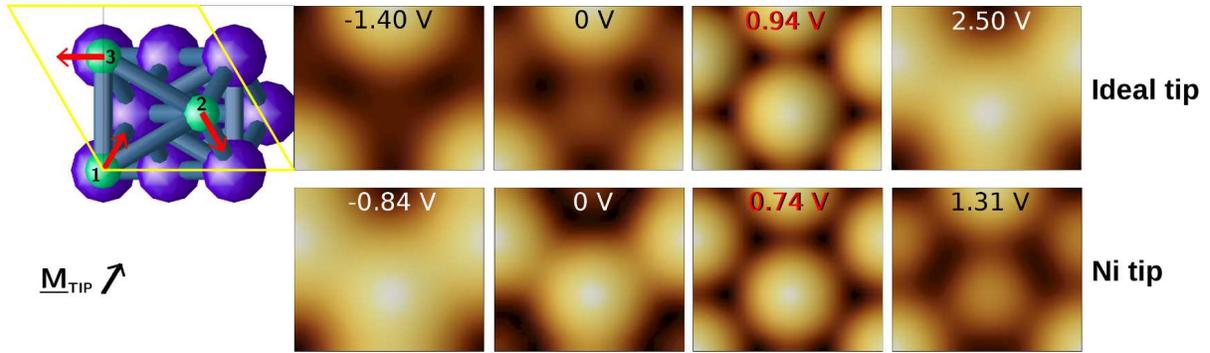}
\caption{\label{Fig4} (Color online) Simulated SP-STM images depending on the bias voltage and the considered tip:
ideal magnetic tip (top row), Ni tip (bottom row). The tip magnetization direction is fixed parallel to the Cr1 magnetic moment,
and is indicated by a vector ($\underline{M}_{TIP}$). The bias values have been chosen corresponding to the local absolute maxima
of the magnetic contrasts and the contrast reversals in Figure \ref{Fig3}.
The surface geometry and the magnetic structure of Cr/Ag(111) as well as the scanning area are also shown.
}
\end{figure*}

In order to better visualize the bias dependent magnetic contrast, we simulated constant current SP-STM images.
Figure \ref{Fig4} shows such images measured with the ideal magnetic tip (top row) and the Ni tip (bottom row) at 0 V,
at the bias voltages corresponding to the contrast reversal (0.94 V and 0.74 V, respectively), and at the voltages
corresponding to the local absolute maxima of the magnetic contrast in both the negative and positive ranges, i.e.,
at -1.40 V and 2.50 V for the ideal tip, and at -0.84 V and 1.31 V using the Ni tip. The surface geometry and the magnetic
structure as well as the scanning area are also shown.
The tip magnetization direction is fixed parallel to the Cr1 magnetic moment. For the zero bias images a temperature of 4.2 K
was considered following Ref.\ \cite{palotas11stm} since there is no current at 0 K.

We find a similar type of magnetic contrast at 0 V and at the negative bias voltages for both tips, respectively, i.e.,
for $V\le 0$, $\Delta z^{Cr1-Cr3}>0$ (Cr1 appears higher than Cr2 and Cr3) for the ideal tip, and $\Delta z^{Cr1-Cr3}<0$
(Cr1 appears lower than Cr2 and Cr3) for the Ni tip, similarly as reported in Figure \ref{Fig3}.
At the corresponding reversal voltages, 0.94 V (ideal tip) and 0.74 V (Ni tip), all Cr atoms appear to be of equal height on the
SP-STM image. Here, the magnetic contrast is lost as the magnetic current is zero. This corresponds to a qualitatively similar
image of performing the STM measurement with a nonmagnetic tip \cite{palotas11stm}. Thus, the magnetic current calculated at a
single point above the surface was indeed able to find the correct bias position of the magnetic contrast reversal for both tips.
Above the reversal voltage, the magnetic contrast is inverted.
We illustrate this by showing the images calculated at 2.50 V for the ideal tip, and at 1.31 V for the Ni tip in Figure \ref{Fig4}.

\begin{figure*}
\includegraphics[width=1.0\columnwidth,angle=0]{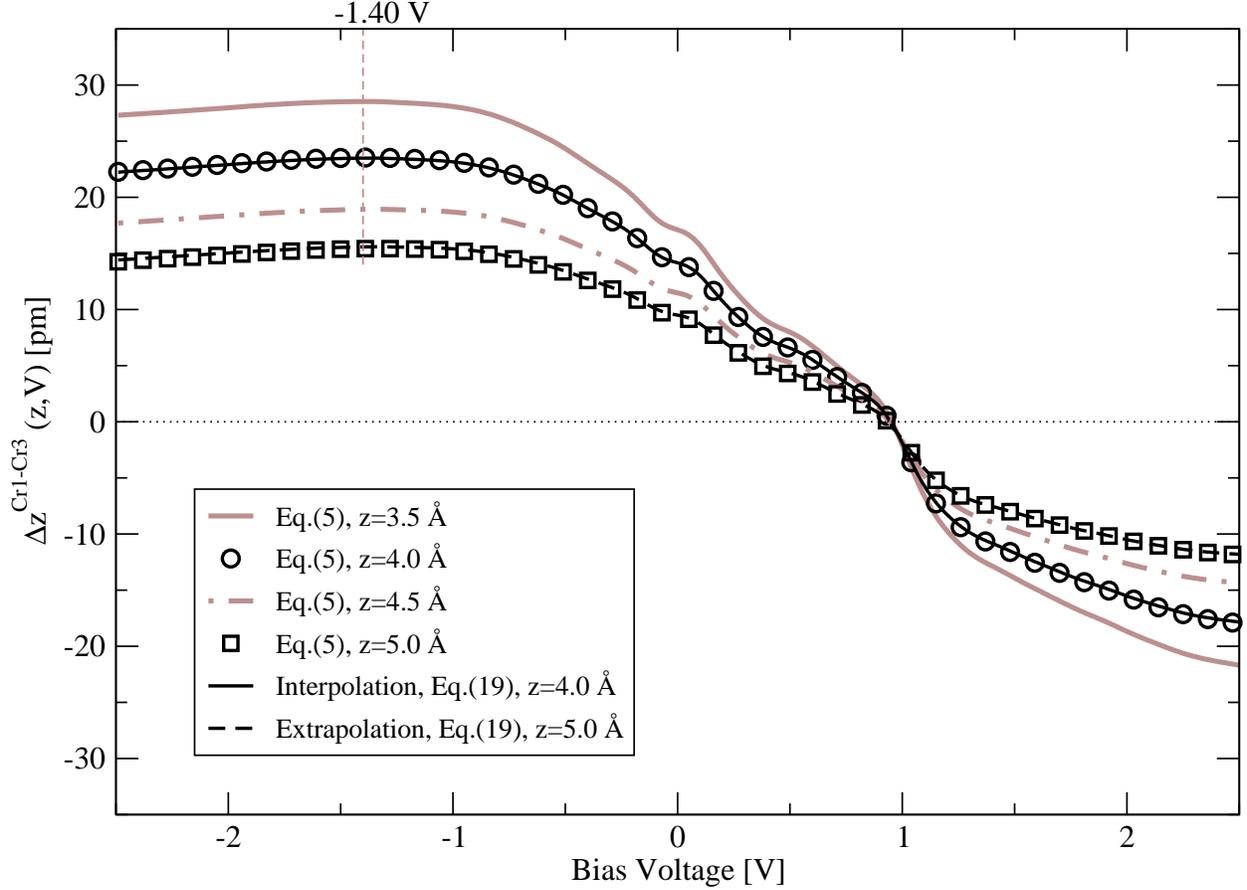}
\caption{\label{Fig5} (Color online) The bias dependent magnetic contrast between Cr1 and Cr3 atoms
$\Delta z^{Cr1-Cr3}(z,V)$ calculated using Eq.(\ref{Eq_DZ}) with the ideal magnetic tip at different tip-sample separations:
$z=3.5$ \AA\;(gray solid line), $z=4.0$ \AA\;(black circles), $z=4.5$ \AA\;(gray dash-dotted line),
and $z=5.0$ \AA\;(black squares).
The tip magnetization direction is fixed parallel to the Cr1 magnetic moment.
Taking the functions colored by gray at $z=3.5$ \AA\;and $z=4.5$ \AA, Eq.(\ref{Eq_corrug4points}) is used to interpolate
the bias dependent magnetic contrast to $z=4.0$ \AA\;(black solid line), and to extrapolate to $z=5.0$ \AA\;(black dashed line).
The absolute contrast maximum of each curve is found at -1.40 V, and is explicitly indicated.
}
\end{figure*}

Let us analyze the effect of the tip-sample distance on the obtained results. Figure \ref{Fig5} shows the bias dependent
magnetic contrast between Cr1 and Cr3 atoms calculated using Eq.(\ref{Eq_DZ}) with the ideal magnetic tip at different tip-sample
separations: $z=3.5$, 4.0, 4.5, and 5.0 \AA. The tip magnetization direction is fixed parallel to the Cr1 magnetic moment.
It is clearly seen that the bias positions of the contrast reversal and the local maxima are practically unaffected
by the tip-sample distance. We tested this for other distances as well, and the bias positions of the local maxima
were found within $\pm 0.02$ V deviation. On the other hand, we find that the absolute contrasts are decreasing with increasing
tip-sample distance. We would like to test whether this decay is exponential. In fact, assuming an exponential decay,
Eq.(\ref{Eq_corrug4points}) is derived in the appendix, which enables the determination of the bias dependent magnetic contrast at
arbitrary tip-sample distances from two contrast functions at fixed heights $z_1$ and $z_2$. This is the third level of
information on the bias dependent magnetic contrast from single point measurements, see section \ref{sec_spec}. Taking the
functions at $z_1=3.5$ and $z_2=4.5$ \AA, Eq.(\ref{Eq_corrug4points}) is used to interpolate the bias dependent magnetic contrast
to $z=4.0$ \AA, and to extrapolate to $z=5.0$ \AA. As the results obtained by the interpolation and the extrapolation agree
quantitatively well with those calculated by using Eq.(\ref{Eq_DZ}), the exponential decay of the absolute magnetic contrast
with respect to the tip-sample distance and, thus, the validity of Eq.(\ref{Eq_corrug4points}) are confirmed within the
presented atom superposition framework. Note that orbital-dependent tunneling effects can modify this finding.

\begin{figure*}
\includegraphics[width=1.0\columnwidth,angle=0]{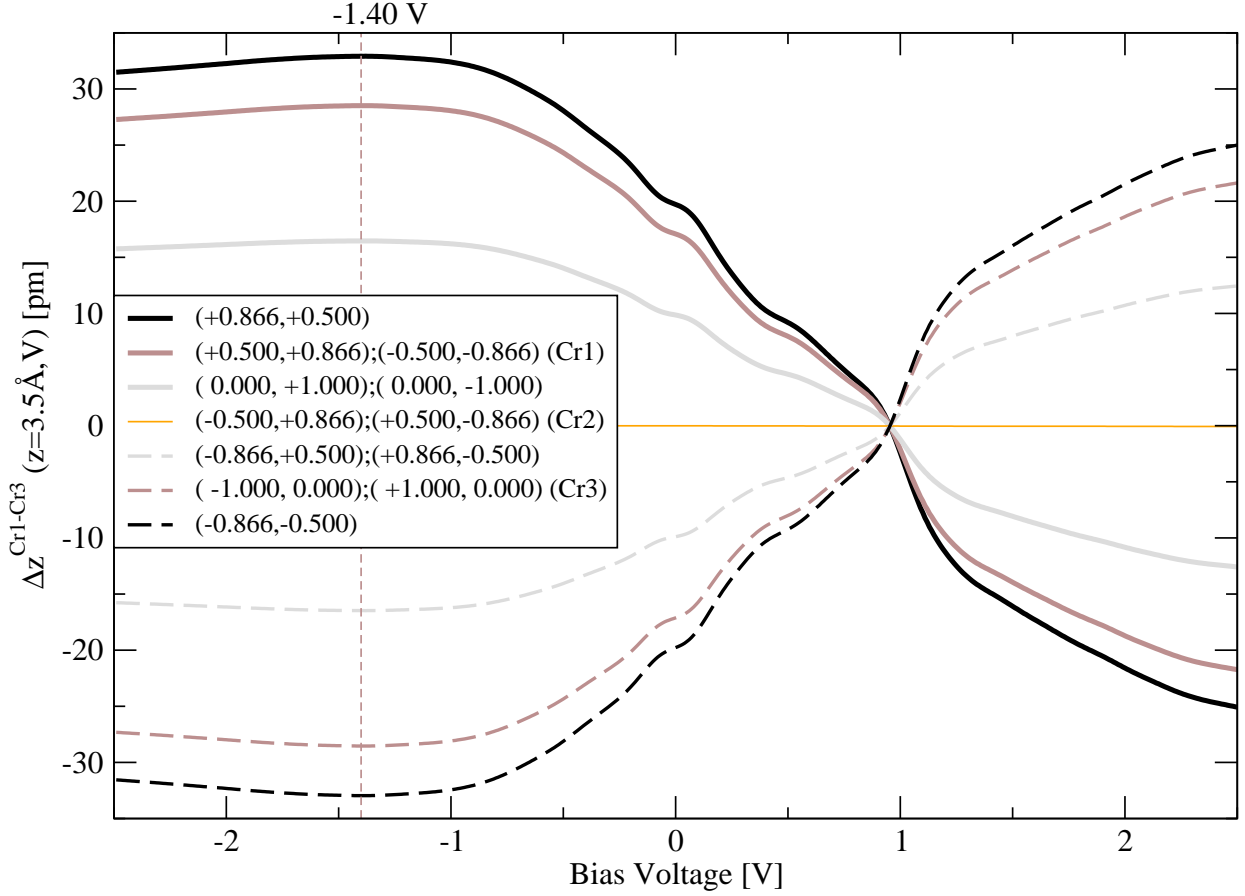}
\caption{\label{Fig6} (Color online) The effect of the tip magnetization orientation on the bias dependent magnetic contrast
between Cr1 and Cr3 atoms $\Delta z^{Cr1-Cr3}(z=3.5$\AA$,V)$ at $z=3.5$ \AA\;tip-sample distance calculated using Eq.(\ref{Eq_DZ})
with the ideal magnetic tip.
The unit vectors of the in-plane tip magnetization orientations are rotated in steps of $30^{\circ}$, and are explicitly shown.
The parallel or antiparallel orientations with respect to the corresponding Cr magnetic moments are given in parentheses.
The absolute contrast maximum of each curve is found at -1.40 V, and is explicitly indicated.
}
\end{figure*}

Finally, let us analyze the effect of the magnetization orientation of the tip on the magnetic contrast.
Figure \ref{Fig6} shows the bias dependent magnetic contrast between Cr1 and Cr3 atoms calculated by Eq.(\ref{Eq_DZ})
at $z=3.5$ \AA\; tip-sample distance using the ideal magnetic tip at twelve in-plane magnetic directions
rotated in steps of $30^{\circ}$. The unit vectors of these directions are explicitly shown for each curve.
From the figure it is clear that the bias positions of the contrast reversal and the local absolute maxima remain unaffected.
The contrast curve indicated by the (Cr1) direction is the same as the corresponding curves in Figure \ref{Fig3} (ideal tip) and 
in Figure \ref{Fig5} ($z=3.5$ \AA). Moreover,
we find that the tip magnetization direction parallel or antiparallel to the Cr2 magnetic moment results in a bias-independent
zero magnetic contrast between Cr1 and Cr3 atoms. This is clear since the magnetic currents above Cr1 and Cr3 atoms are equal
in this case, similarly as the magnetic currents above Cr2 and Cr3 atoms are equal [see Eq.(\ref{Eq_ICr3})] in Figure \ref{Fig2}
when the tip magnetization is parallel to the Cr1 moment.
Apart from this, Figure \ref{Fig6} shows that the magnetic contrast can be even more enhanced if we change the tip magnetization
direction from parallel to the Cr1 moment (+0.500,+0.866) to the (+0.866,+0.500) direction. This latter direction provides the
largest achievable positive magnetic contrast between Cr1 and Cr3 atoms at -1.40 V bias. Turning the tip magnetization orientation
to the opposite (-0.866,-0.500) direction, a reversed curve is obtained. This direction of the tip magnetization enables to reach
the largest negative magnetic contrast between Cr1 and Cr3 atoms at -1.40 V bias.
This finding suggests the possibility of tuning the magnetic contrast not only by changing the bias voltage but also by
changing the tip magnetization direction.

Furthermore, each tip magnetization orientation can be characterized by an effective tip spin polarization factor, which gives the
ratio of their contrast curves to a curve corresponding to a prescribed tip magnetization orientation. If we fix the tip
magnetization to the (+0.866,+0.500) direction, effective tip spin polarization factors of 1, 0.866, 0.5, 0, -0.5, -0.866, and -1
can be assigned to the contrast curves listed in the legend of Figure \ref{Fig6}, respectively. This means, e.g., that the
same contrast curve is obtained for a combination of (+0.866,+0.500) tip magnetization direction and $P_T=0.866$ as for the
combination of (+0.500,+0.866) tip magnetization direction and $P_T=1$, the curve denoted by (Cr1) in Figure \ref{Fig6}.

\section{Conclusions}
\label{sec_concl}

In this work we presented a theoretical description of the contrast, i.e., the apparent height difference between two lateral
surface positions on constant current SP-STM images. We introduced two magnetic contrast formulas, and proposed a procedure to
predict different levels of information on the bias voltage dependent magnetic contrast from single point tunneling current or
differential conductance measurements, without the need of scanning large areas of the surface.
Depending on the number of single point measurements, the bias positions of magnetic contrast reversals and of the maximally
achievable magnetic contrast can be determined. Furthermore, we proposed that the bias dependent magnetic contrast between two
surface sites for arbitrary tip-sample separations can be obtained from tunneling measurements at four well-defined tip positions.
We validated our proposals by means of numerical simulations on the complex magnetic surface Cr/Ag(111) employing the atom
superposition approach within the independent orbital approximation based on first principles electronic structure calculations.
Comparing the bias dependence of the predicted magnetic contrasts to that of extracted from constant current SP-STM images we
found excellent agreement. Our results suggest that both proposed formulas are needed for a quantitative determination of the
bias dependent magnetic contrast.
Analyzing the tip-sample distance dependence of the contrast, we found that the bias positions of the contrast reversal and the
local maxima are practically unaffected, and the magnitude of the magnetic contrast is exponentially decaying with respect to
the tip-sample distance. Moreover, we showed evidence that the tip electronic structure and magnetic orientation have a
major effect on the magnetic contrast.
Our theoretical prediction is expected to inspire experimentalists to considerably reduce measurement efforts for determining the
bias dependent magnetic contrast on magnetic surfaces.

\section*{Acknowledgments}

The author thanks L\'aszl\'o Szunyogh for useful comments on the manuscript. Financial support of the Magyary Foundation,
EEA and Norway Grants, the Hungarian Scientific Research Fund (OTKA PD83353, K77771), the Bolyai Research Grant of the
Hungarian Academy of Sciences, and the New Sz\'echenyi Plan of Hungary (Project ID: T\'AMOP-4.2.2.B-10/1--2010-0009)
is gratefully acknowledged. Furthermore, partial usage of the computing facilities of the Wigner Research Centre for Physics
and the BME HPC Cluster is kindly acknowledged.

\appendix
\section{Derivation of Eq.(\ref{Eq_corrug4points})}
\label{sec_app}

Let us assume that the magnitude of the magnetic contrast decays exponentially with increasing tip-sample distance,
\begin{equation}
|\Delta z^{AB}(z,V)|=\Delta z^{AB}_0(V)e^{-2\kappa_{\Delta z}^{AB}(V)z}.
\end{equation}
Expressing the natural logarithm of the magnitude of the contrast at two different tip-sample distances $z_1$ and $z_2$,
we can write:
\begin{eqnarray}
\label{Eq_dz1}
\ln |\Delta z^{AB}(z_1,V)|=\ln\Delta z^{AB}_0(V)-2\kappa_{\Delta z}^{AB}(V)z_1,\\
\ln |\Delta z^{AB}(z_2,V)|=\ln\Delta z^{AB}_0(V)-2\kappa_{\Delta z}^{AB}(V)z_2.
\label{Eq_dz2}
\end{eqnarray}
Subtracting Eq.(\ref{Eq_dz2}) from Eq.(\ref{Eq_dz1}), the inverse decay length of the contrast can be given as
\begin{equation}
\label{Eq_kappadz}
\kappa_{\Delta z}^{AB}(V)=\frac{\ln |\Delta z^{AB}(z_1,V)|-\ln |\Delta z^{AB}(z_2,V)|}{2(z_2-z_1)}.
\end{equation}
On the other hand, by adding Eq.(\ref{Eq_dz2}) to Eq.(\ref{Eq_dz1}), $\ln\Delta z^{AB}_0(V)$ can be obtained:
\begin{eqnarray}
2\ln\Delta z^{AB}_0(V)&=&\ln |\Delta z^{AB}(z_1,V)|+\ln |\Delta z^{AB}(z_2,V)|+2\kappa_{\Delta z}^{AB}(V)(z_1+z_2)\nonumber\\
&=&\left(1+\frac{z_1+z_2}{z_2-z_1}\right)\ln |\Delta z^{AB}(z_1,V)|+\left(1-\frac{z_1+z_2}{z_2-z_1}\right)\ln |\Delta z^{AB}(z_2,V)|\nonumber\\
&=&2\frac{z_2}{z_2-z_1}\ln |\Delta z^{AB}(z_1,V)|-2\frac{z_1}{z_2-z_1}\ln |\Delta z^{AB}(z_2,V)|.
\end{eqnarray}
Thus, the theoretical magnetic contrast at $z=0$ can be written as
\begin{equation}
\Delta z^{AB}_0(V)=\frac{|\Delta z^{AB}(z_1,V)|^{\frac{z_2}{z_2-z_1}}}{|\Delta z^{AB}(z_2,V)|^{\frac{z_1}{z_2-z_1}}}.
\end{equation}
Using the above quantities, the magnitude of the magnetic contrast at an arbitrary tip-sample distance $z$ can be given as
\begin{eqnarray}
|\Delta z^{AB}(z,V)|&=&\Delta z^{AB}_0(V)e^{-2\kappa_{\Delta z}^{AB}(V)z}\nonumber\\
&=&|\Delta z^{AB}(z_1,V)|^{\frac{z_2}{z_2-z_1}}\times |\Delta z^{AB}(z_2,V)|^{\frac{-z_1}{z_2-z_1}}
\times e^{\frac{z}{z_2-z_1}\left(\ln |\Delta z^{AB}(z_2,V)|-\ln |\Delta z^{AB}(z_1,V)|\right)}\nonumber\\
&=&|\Delta z^{AB}(z_1,V)|^{\frac{z_2-z}{z_2-z_1}}\times |\Delta z^{AB}(z_2,V)|^{\frac{z-z_1}{z_2-z_1}}.
\end{eqnarray}
Taking the sign of the magnetic contrast into account, we arrive at the formula reported in Eq.(\ref{Eq_corrug4points}).

\begin{table}
{\scriptsize
\begin{tabular}{ccccc}\hhline{=====}
Information on the         &Tip      &\multicolumn{2}{c}{Measured quantities}&Derived quantities                  \\
magnetic contrast          &positions&Differential conductance&Current       &                                    \\ \hline
Bias position of           &$(A,z_1)$&$dI_P^A/dV(z_1,V)$   &$I_P^A(z_1,V)$   &                                    \\
contrast reversal          &         &$dI_{AP}^A/dV(z_1,V)$&$I_{AP}^A(z_1,V)$&$I_{MAGN}^A(z_1,V)$                 \\
(1st level)                &         &                     &                 &                                    \\ \hline
Bias dependent contrast    &$(B,z_1)$&$dI_P^B/dV(z_1,V)$   &$I_P^B(z_1,V)$   &                                    \\
between atoms $A$ and $B$  &         &$dI_{AP}^B/dV(z_1,V)$&$I_{AP}^B(z_1,V)$&$\mathbf{I_{MAGN}^B(z_1,V)}$        \\
at the fixed height $z_1$: &$(A,z_1)$&$dI_P^A/dV(z_1,V)$   &$I_P^A(z_1,V)$   &$\mathbf{I_{MAGN}^A(z_1,V)}$        \\
$\mathbf{\Delta z^{AB}_I(z_1,V)}$ using {\bf Eq.(\ref{Eq_DZ})}&&$dI_{AP}^A/dV(z_1,V)$&$I_{AP}^A(z_1,V)$&$\mathbf{I_{TOPO}^A(z_1,V)}$\\
(2nd level)                &$(A,z_2)$&$dI_P^A/dV(z_2,V)$   &$I_P^A(z_2,V)$   &$I_{TOPO}^A(z_2,V)$                 \\
                           &         &$dI_{AP}^A/dV(z_2,V)$&$I_{AP}^A(z_2,V)$&$\mathbf{\kappa_{TOPO}^A(V)}$       \\ \hline
Bias dependent contrast    &$(A,z_1)$&$dI_P^A/dV(z_1,V)$   &$I_P^A(z_1,V)$   &                                    \\
between atoms $A$ and $B$  &         &$dI_{AP}^A/dV(z_1,V)$&$I_{AP}^A(z_1,V)$&$\mathbf{I_{MAGN}^A(z_1,V)}$        \\
at the fixed height $z_1$: &$(B,z_1)$&$dI_P^B/dV(z_1,V)$   &$I_P^B(z_1,V)$   &$\mathbf{I_{MAGN}^B(z_1,V)}$        \\
$\mathbf{\Delta z^{AB}_{II}(z_1,V)}$ using {\bf Eq.(\ref{Eq_DZ-TOT})}&&$dI_{AP}^B/dV(z_1,V)$&$I_{AP}^B(z_1,V)$&$\mathbf{I_{TOT}^{av.AB}(z_1,V)}$\\
(2nd level)                &$(A,z_2)$&$dI_P^A/dV(z_2,V)$   &$I_P^A(z_2,V)$   &$I_{TOT}^{av.AB}(z_2,V)$            \\
                           &$(B,z_2)$&$dI_P^B/dV(z_2,V)$   &$I_P^B(z_2,V)$   &$\mathbf{\kappa_{TOT}^{av.AB}(V)}$  \\ \hline
Bias dependent contrast    &$(B,z_1)$&$dI_P^B/dV(z_1,V)$   &$I_P^B(z_1,V)$   &$I_{MAGN}^B(z_1,V)$                 \\
between atoms $A$ and $B$  &         &$dI_{AP}^B/dV(z_1,V)$&$I_{AP}^B(z_1,V)$&$I_{MAGN}^A(z_1,V)$                 \\
at an arbitrary height $z$:&$(A,z_1)$&$dI_P^A/dV(z_1,V)$   &$I_P^A(z_1,V)$   &$I_{TOPO}^A(z_1,V)$                 \\
$\mathbf{\Delta z^{AB}(z,V)}$ using {\bf Eq.(\ref{Eq_corrug4points})}&&$dI_{AP}^A/dV(z_1,V)$&$I_{AP}^A(z_1,V)$&$I_{TOPO}^A(z_2,V)$\\
based on {\bf Eq.(\ref{Eq_DZ})}&$(A,z_2)$&$dI_P^A/dV(z_2,V)$&$I_P^A(z_2,V)$  &$\kappa_{TOPO}^A(V)$                \\
(3rd level)                &         &$dI_{AP}^A/dV(z_2,V)$&$I_{AP}^A(z_2,V)$&$I_{MAGN}^A(z_2,V)$                 \\
                           &$(B,z_2)$&$dI_P^B/dV(z_2,V)$   &$I_P^B(z_2,V)$   &$I_{MAGN}^B(z_2,V)$                 \\
                           &         &$dI_{AP}^B/dV(z_2,V)$&$I_{AP}^B(z_2,V)$&$\mathbf{\Delta z^{AB}_I(z_1,V)}$   \\
                           &         &                     &                 &$\mathbf{\Delta z^{AB}_I(z_2,V)}$   \\ \hline
Bias dependent contrast    &$(A,z_1)$&$dI_P^A/dV(z_1,V)$   &$I_P^A(z_1,V)$   &$I_{MAGN}^A(z_1,V)$                 \\
between atoms $A$ and $B$  &         &$dI_{AP}^A/dV(z_1,V)$&$I_{AP}^A(z_1,V)$&$I_{MAGN}^B(z_1,V)$                 \\
at an arbitrary height $z$:&$(B,z_1)$&$dI_P^B/dV(z_1,V)$   &$I_P^B(z_1,V)$   &$I_{TOT}^{av.AB}(z_1,V)$            \\
$\mathbf{\Delta z^{AB}(z,V)}$ using {\bf Eq.(\ref{Eq_corrug4points})}&&$dI_{AP}^B/dV(z_1,V)$&$I_{AP}^B(z_1,V)$&$I_{TOT}^{av.AB}(z_2,V)$\\
based on {\bf Eq.(\ref{Eq_DZ-TOT})}&$(A,z_2)$&$dI_P^A/dV(z_2,V)$&$I_P^A(z_2,V)$&$\kappa_{TOT}^{av.AB}(V)$         \\
(3rd level)                &         &$dI_{AP}^A/dV(z_2,V)$&$I_{AP}^A(z_2,V)$&$I_{MAGN}^A(z_2,V)$                 \\
                           &$(B,z_2)$&$dI_P^B/dV(z_2,V)$   &$I_P^B(z_2,V)$   &$I_{MAGN}^B(z_2,V)$                 \\
                           &         &$dI_{AP}^B/dV(z_2,V)$&$I_{AP}^B(z_2,V)$&$\mathbf{\Delta z^{AB}_{II}(z_1,V)}$\\
                           &         &                     &                 &$\mathbf{\Delta z^{AB}_{II}(z_2,V)}$\\ 
\hhline{=====}
\end{tabular}
}
\caption{\label{Table_method}
The summary of the procedure to obtain different levels of information on the bias dependent magnetic contrast from single point
differential conductance or current measurements at the indicated tip positions with oppositely magnetized $P$ and $AP$ tips.
The derived quantities are necessary for the magnetic contrast formulas, Eq.(\ref{Eq_DZ}), Eq.(\ref{Eq_DZ-TOT}), and
Eq.(\ref{Eq_corrug4points}), and the direct ingredients are given in boldface, respectively.
}
\end{table}

\newpage

\end{document}